\documentclass[runningheads]{llncs}
\usepackage[T1]{fontenc}
\usepackage[colorlinks=true,linkcolor=blue,citecolor=blue,urlcolor=blue,]{hyperref}
 
\usepackage[utf8]{inputenc} % allow utf-8 input
\usepackage[T1]{fontenc}    % use 8-bit T1 fonts
\usepackage{hyperref}       % hyperlinks
\usepackage{url}            % simple URL typesetting
\usepackage{booktabs}       % professional-quality tables
\usepackage{mathtools,amssymb}
\usepackage{amsfonts}       % blackboard math symbols
\usepackage{nicefrac}       % compact symbols for 1/2, etc.
\usepackage{microtype}      % microtypography
\usepackage{array,colortbl}
\usepackage[dvipsnames]{xcolor}
\usepackage{algorithm,algorithmicx,algpseudocode}
\usepackage{graphbox}
\usepackage{placeins}
\usepackage{wrapfig}
\usepackage{etoolbox}
\usepackage{multirow}
\usepackage{graphicx}

\newcommand{\grad}{\nabla}

\newcommand{\bI}{\mathbf{I}}

\newcommand{\bzero}{\mathbf{0}}

\newcommand{\bz}{\mathbf{z}}

\newcommand{\bepsilon}{{\boldsymbol{\epsilon}}}

\begin{document}
\title{Learning Enhancement From Degradation: A Diffusion Model For Fundus Image Enhancement}
\vspace{-40mm}
%\thanks{Supported by organization x.}}
%
\titlerunning{Learning enhancement from degradation}
%\author{First Author\inst{1}\orcidID{0000-1111-2222-3333} \and
%Second Author\inst{2,3}\orcidID{1111-2222-3333-4444} \and
%Third Author\inst{3}\orcidID{2222--3333-4444-5555}}
\author{Pujin Cheng$^{1}$, Li Lin$^{1, 2}$, Yijin Huang $^{1,3}$, Huaqing He$^{1}$,Wenhan Luo$^{4}$, Xiaoying Tang$^{1}$\thanks{Corresponding author: Dr. Xiaoying Tang (\url{tangxy@sustech.edu.cn}).}}
\authorrunning{Cheng et al.}
\institute{$^{1}$Department of Electronic and Electrical Engineering, \\ Southern University of Science and Technology, Shenzhen, China
\\
$^{2}$ Department of Electrical and Electronic Engineering,\\ The University of Hong Kong, Hong Kong SAR, China
\\
$^{3}$School of Biomedical Engineering,\\
University of British Columbia, Vancouver, British Columbia, Canada
\\
$^{4}$Sun Yat-sen University, Shenzhen, China
}
\vspace{-40mm}
%\institute{Princeton University, Princeton NJ 08544, USA \and
%Springer Heidelberg, Tiergartenstr. 17, 69121 Heidelberg, Germany
\maketitle              % typeset the header of the contribution
\begin{abstract}
The quality of a fundus image can be compromised by numerous factors, many of which are challenging to be appropriately and mathematically modeled. In this paper, we introduce a novel diffusion model based framework, named Learning Enhancement from Degradation (LED), for enhancing fundus images. Specifically, we first adopt a data-driven degradation framework to learn degradation mappings from unpaired high-quality to low-quality images. We then apply a conditional diffusion model to learn the inverse enhancement process in a paired manner. The proposed LED is able to output enhancement results that maintain clinically important features with better clarity. Moreover, in the inference phase, LED can be easily and effectively integrated with any existing fundus image enhancement framework. We evaluate the proposed LED on several downstream tasks with respect to various clinically-relevant metrics, successfully demonstrating its superiority over existing state-of-the-art methods both quantitatively and qualitatively. The source code is available at \url{https://github.com/QtacierP/LED}.

\end{abstract}
\section{Introduction}
The quality of a fundus image is crucial for the diagnosis and treatment of various ophthalamic diseases. However, many factors may compromise the quality of a fundus image, including glare, motion blur, and poor illumination~\cite{human,clahe,quality-survey}. Although many methods have been proposed to enhance fundus images~\cite{sgrif,dai,de-clahe}, it is still a challenging task due to the lack of appropriate mathematical models to characterize the complex degradation processes. Recent researches mainly focus on learning enhancement in data-driven fashions.

\begin{figure}[t]
  \centering
  \includegraphics[width=1.0\textwidth]{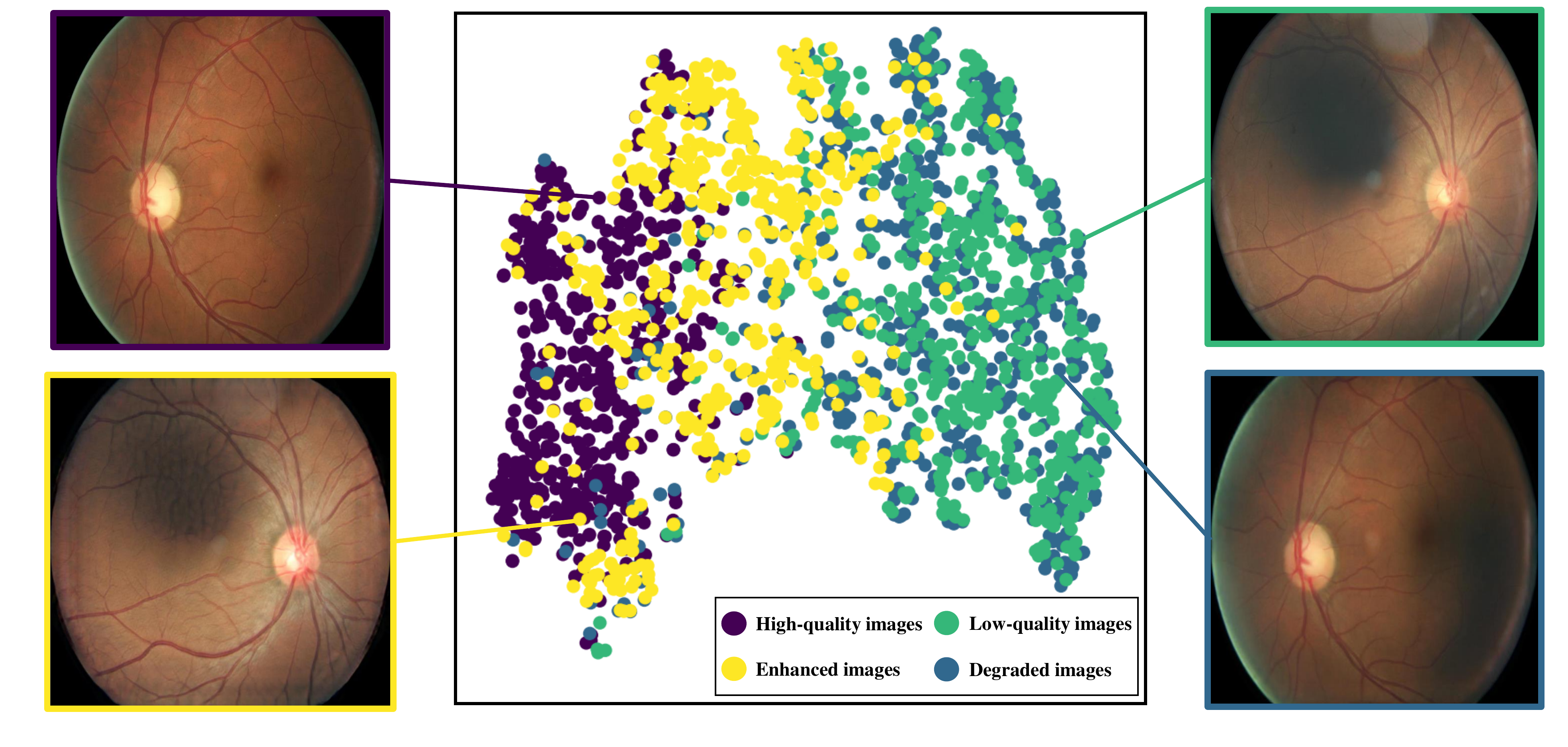}
  \vspace{-6mm}
  \caption{t-SNE~\cite{tsne} visualization of realistic and synthetic images' features extracted from a pre-trained quality assessment model. The low-quality and high-quality images are real-world fundus images, while the degraded and enhanced images are generated by a well-trained CycleGAN consisting of a degradation network and an enhancement network.}
  \label{fig:degradation}
  \vspace{-8mm}
  \end{figure}

To perform data-driven image enhancement, two primary types of approaches have been proposed: degradation modeling based enhancement and unpaired image translation based enhancement. Degradation modeling based methods explicitly model the degradation process to restore degraded images to their original quality, whereas unpaired image translation based methods map images to a visually more appealing domain with no need for modeling the degradation process. Shen et al. develop a degradation framework based on the imaging principles of fundus cameras, which is then used to design a correction network named CofeNet~\cite{CoFoNet}. Built on this same degradation pipeline, Liu et al. propose the Pyramid Constraint Network (PCENet) to enhance clinically-relevant representation~\cite{PCENet}. Li et al. present the Structure-Consistent Restoration Network (SCRNet) for cataract fundus images, which lays its foundation on the consistency of high-frequency components~\cite{SCRNet}. On the other hand, unpaired image translation based enhancement methods rely on adversarial training and do not require explicit modeling of image degradation. For example, Zhao et al. introduce a CycleGAN-based~\cite{cycle} network for deblurring fundus images in an unpaired manner~\cite{Zhao}.

The performance of degradation modeling based enhancement methods is mainly limited by the complexity of the degradation processes, often resulting in poor performance on real-world low-quality fundus images. On the other hand, unpaired image translation based enhancement methods are trained without any supervision, which may lead to information modification and unrealistic results. To address these issues, Cheng et al. propose a semi-supervised framework named I-SECRET that combines the advantages of both approaches~\cite{I-SECRET}. Nevertheless, it still requires a well-designed degradation model to guide the training process and may fail to preserve clinically-relevant details.  Li et al. introduce an unsupervised domain adaptation framework into ArcNet~\cite{ArcNet}, but that approach may still fail on low-quality images caused by out-of-distribution (OOD) factors.

\begin{figure}[t]
  \includegraphics[width=\textwidth]{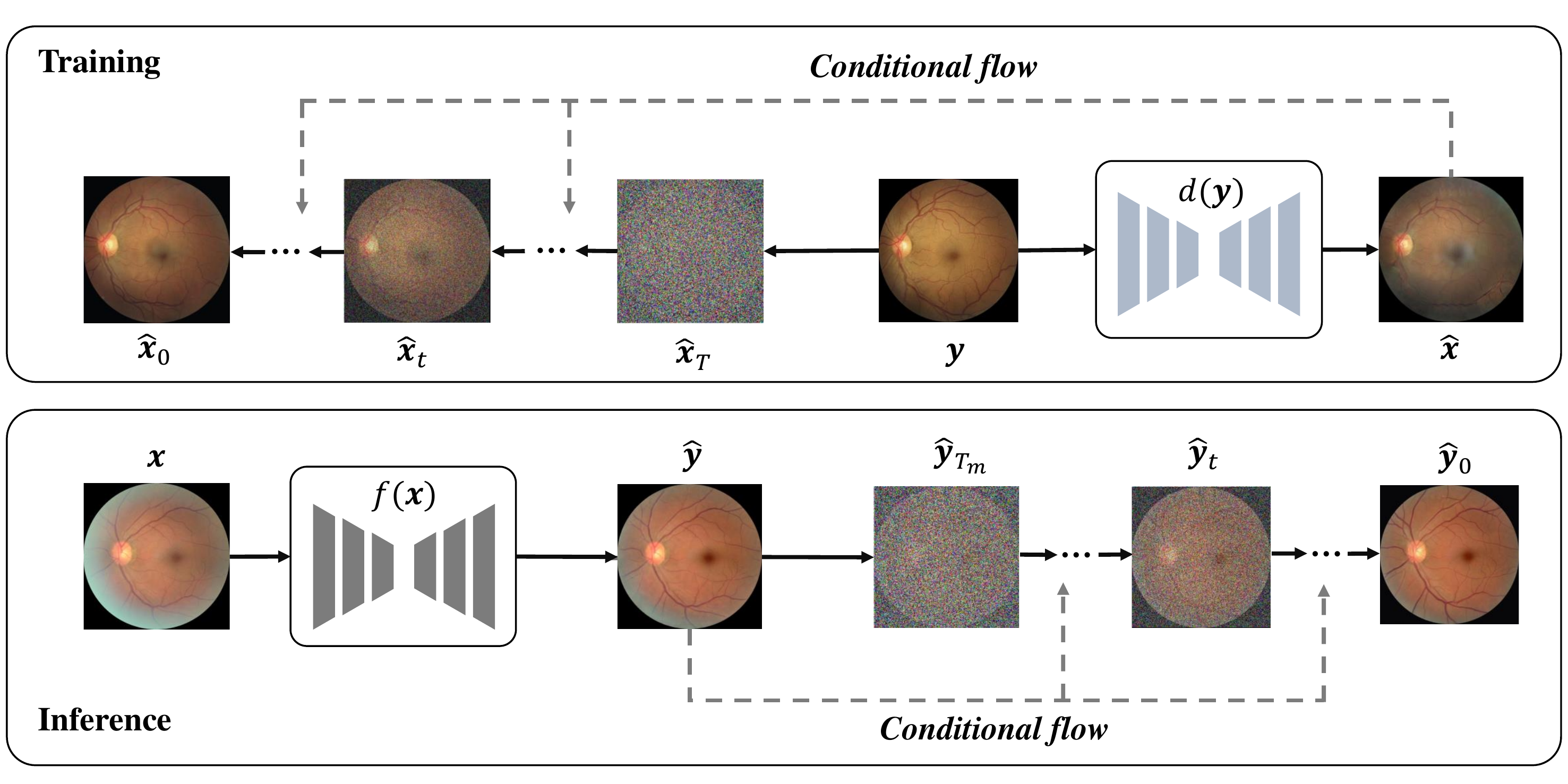}
  \vspace{-8mm}
  \caption{The overall framework of the proposed LED. In the training phase, a high-quality image $\boldsymbol{y}$ is degraded into $\hat{\boldsymbol{x}}$ through a pre-trained data-driven degradation model $d\boldsymbol{(y})$, and the diffusion model aims to restore the degraded image. In the inference phase, given a realistic low-quality image $\boldsymbol{x}$, a coarse enhancement network $f(\boldsymbol{x})$ firstly enhances it to $\boldsymbol{\hat{y}}$, and the diffusion model refines the coarse enhancement $\boldsymbol{\hat{y}}$ into the final enhancement $\boldsymbol{\hat{y}}_0$.}\label{fig:overall}
  \vspace{-6mm}
  \end{figure}

In this paper, we propose a novel fundus image enhancement framework that utilizes a diffusion model to Learn the Enhancement from a data-driven Degradation model (LED), motivated by the observation that the distance between data-driven degraded images and realistic low-quality images is much smaller than that between data-driven enhanced images and high-quality images, as shown in Fig. \ref{fig:degradation}. Specifically, LED overcomes the limitations of traditional enhancement methods that often rely on manually-designed degradation models; those models typically cannot fully capture the complexity of real-world fundus image degradation. Our employed diffusion model can also effectively avoid OOD enhancement by modeling the distribution of high-quality images. Unlike existing diffusion model based image reconstruction methods that sample directly from the noisy input image~\cite{anoddpm,abnormal,diffusion-survey,towards}, our proposed LED framework embeds the high-quality image into a Gaussian distribution with an additional input of the paired non-noisy low-quality image, providing smooth and continuous enhancement while preserving clinically-relevant details and attenuating information modification.  The main contributions of this paper are four-fold: (1) To the best of our knowledge, LED is the first diffusion model based method for fundus image enhancement. (2) Our proposed diffusion process delivers realistic and reliable enhancement that does not rely on any well-designed degradation model. (3) Our framework is flexible and can be easily and effectively integrated into any existing fundus image enhancement framework. (4) Comprehensive experiments on several downstream tasks with respect to clinically-relevant metrics demonstrate the superiority of LED over existing state-of-the-art (SOTA) methods.

\section{Methods}
The overall framework is shown in Fig. \ref{fig:overall}.  Given a training set $\mathcal{X}=\{\boldsymbol{x}_i\}_{i=1}^N$, $\mathcal{Y}=\{\boldsymbol{y}_i\}_{i=1}^M$, where $\boldsymbol{x}_i$ and $\boldsymbol{y}_i$ are respectively the low-quality and high-quality images, we first learn a degradation mapping $d(\boldsymbol{y})$ from $\mathcal{Y}$ to $\mathcal{X}$ with an unpaired image translation framework. Then, we apply a conditional diffusion model to learn the process of enhancing the degraded images to the original high-quality images. In the inference phase, we can integrate LED with any existing fundus image enhancement framework $f(\boldsymbol{x})$. In the following paragraphs, we firstly introduce diffusion models and then present our proposed LED in detail.

\subsection{Diffusion Models}
We present a brief introduction to diffusion models in this section. For more details, please refer to~\cite{DDPM}. Diffusion models are a class of generative models that can be used to generate samples from Gaussian noises. In general, diffusion models consist of two parts: a reverse process and a forward process. 

Given a sample $\boldsymbol{x}_0$ and its corresponding distribution $q(\boldsymbol{x}_0)$, the forward process is used to generate a sequence of samples $\{\boldsymbol{x}_t\}_{t=1}^T$ based on $q(\boldsymbol{x}_0)$, the samples of which shall be corrupted by Gaussian noises. This process can be formulated as
\begin{equation}
    q\left(\boldsymbol{x}_t \mid \boldsymbol{x}_{t-1}\right)=\mathcal{N}\left(\boldsymbol{x}_t ; \sqrt{1-\beta_t} \boldsymbol{x}_{t-1}, \beta_t \boldsymbol{I}\right), \ \ \ t=1,2,3,\cdots,T,
\end{equation}
where $[\beta_1, \ldots, \beta_T]$ denotes a pre-defined or learned variance schedule. Sampling from $q(\boldsymbol{x}_0)$ at time step $t$ is denoted as
\begin{equation}
    q\left(\boldsymbol{x}_t \mid \boldsymbol{x}_0\right)=\mathcal{N}\left(\boldsymbol{x}_t ; \sqrt{\alpha_t} \boldsymbol{x}_0,\left(1-\alpha_t\right) \boldsymbol{I}\right),
\label{eq:2}
\end{equation}
where $\alpha_t=\prod_{i=1}^t\left(1-\beta_i\right)$. The goal of a diffusion model is to learn the reverse process $p(\boldsymbol{x}_0|\boldsymbol{x}_T)$. DDPM~\cite{DDPM} points out that the reverse process $p\left(\boldsymbol{x}_{t-1} \mid \boldsymbol{x}_t\right)$ can be approximated using the following equation
\begin{equation}
    p_\theta\left(\boldsymbol{x}_{t-1} \mid \boldsymbol{x}_t\right):=\mathcal{N}\left(\boldsymbol{x}_{t-1} ; \boldsymbol{\mu}_\theta\left(\boldsymbol{x}_t, t\right), \tilde{\beta}_t \boldsymbol{I}\right),
\end{equation}
where $\tilde{\beta}_t=\frac{1-\alpha_{t-1}}{1-\alpha_t} \beta_t$ and $\theta$ represents the learnable parameters. In this way, the reverse process, which aims to denoise the corrupted samples, is formulated as
\begin{equation}
    \boldsymbol{\mu}_\theta\left(\boldsymbol{x}_t, t\right)=\frac{1}{\sqrt{1-\beta_t}}\left(\boldsymbol{x}_t-\frac{\beta_t}{\sqrt{1-\alpha_t}} \boldsymbol{\epsilon}_\theta\left(\boldsymbol{x}_t, t\right)\right),
\end{equation}
where $\boldsymbol{\epsilon}_\theta$ is the noise estimation network. We further reformulate Eq.\eqref{eq:2} as $\boldsymbol{x}_t\left(\boldsymbol{x}_0, \boldsymbol{\epsilon}\right)=\sqrt{\alpha_t} \boldsymbol{x}_0+\sqrt{1-\alpha_t} \boldsymbol{\epsilon}$ using a reparameterization trick, where $\epsilon \sim \mathcal{N}(\boldsymbol{0}, \boldsymbol{I})$. The objective function of DDPM is
\begin{equation}
  \mathcal{L}_\theta(\boldsymbol{x}_0, t) =  \mathbb{E}_{\boldsymbol{x}_0, \boldsymbol{\epsilon}, t}\left[\left\|\boldsymbol{\epsilon}-\boldsymbol{\epsilon}_\theta\left(\sqrt{\alpha_t} \boldsymbol{x}_0+\sqrt{1-\alpha_t} \boldsymbol{\epsilon}, t\right)\right\|^2\right].
\end{equation}
Practicably, DDPM randomly samples $t$ from $[0, T]$ to minimize the objective function, where $T$ is a hyper-parameter denoting the total number of time steps.

\subsection{Proposed Framework}
Based on DDPM, we design a diffusion model based framework for fundus image enhancement. The key idea is to identify a mapping from a low-quality image of interest to its high-quality version using the reverse process. With an unconditional diffusion model, structural details in the image may nevertheless be modified during the denoising process. To address this issue, we introduce a diffusion model that conditions on the low-quality image, which aims to minimize
\begin{equation}
    \mathcal{L}_\theta(\boldsymbol{x}, \boldsymbol{y}, t) =  \mathbb{E}_{\boldsymbol{x}, \boldsymbol{y}, \boldsymbol{\epsilon}, t}\left[\left\|\boldsymbol{\epsilon}-\boldsymbol{\epsilon}_\theta\left(\sqrt{\alpha_t} \boldsymbol{y} +\sqrt{1-\alpha_t} \boldsymbol{\epsilon}, t, \boldsymbol{x}\right)\right\|^2\right],
\end{equation}
where $\boldsymbol{x}$ is the low-quality image and $\boldsymbol{y}$ is the paired high-quality image. However, it is difficult or even infeasible to obtain the paired high-quality image. To solve this problem, we employ a data-driven degradation model $d(\boldsymbol{y})$ to first generate a paired low-quality image $\boldsymbol{\hat{x}}$ from the high-quality image $\boldsymbol{y}$. This model can be easily obtained from any existing unpaired image translation method. Based on the synthesized low-quality image, we can then train the diffusion model to learn the reverse process $\mathcal{L}_\theta(\boldsymbol{\hat{x}}, \boldsymbol{y}, t)$.

In the inference phase, the high-quality image $\boldsymbol{y}$ is not available. Therefore, we add noises to the low-quality image $\boldsymbol{x}$ itself instead of $\boldsymbol{y}$, under the assumption that the high-frequency components of the noisy low-quality image $\boldsymbol{x}_t$ and the noisy high-quality image $\boldsymbol{y}_t$ shall be similar when $t$ is large. 

Since the condition $\boldsymbol{\hat{x}}$ of the diffusion model is acquired from a data-driven degradation method, it can be theoretically replaced by any artificially generated fundus image. As such, we further propose a coarse-to-fine enhancement strategy in the inference stage. Given any existing image enhancement method $f(\boldsymbol{x})$, we first generate a coarse high-quality image $\boldsymbol{\hat{y}}$ from the low-quality image $\boldsymbol{x}$ through $f(\boldsymbol{x})$, and then employ the proposed LED to refine $\boldsymbol{\hat{y}}$ and produce a final high-quality result $\boldsymbol{\hat{y}}_0$. The diffusion process explicitly models the target high-qualify image distribution, which enables the model to generate more realistic high-quality images by preserving high-frequency details and attenuating structural modifications that may nevertheless occur with a sole application of an existing enhancement method. Algorithm~\ref{alg:training} and Algorithm~\ref{alg:inference} respectively show the training and inference processes of LED.

\algrenewcommand\algorithmicindent{0.5em}%
\begin{figure}[t]
\begin{minipage}[t]{0.495\textwidth}
\begin{algorithm}[H]
  \caption{Training} \label{alg:training}
  \small
  \begin{algorithmic}[1]
    \Repeat
      \State $\boldsymbol{y} \sim q(\boldsymbol{y})$
      \State $\boldsymbol{\hat{x}} = d(\boldsymbol{y})$
      \State $t \sim \mathrm{Uniform}(\{1, \dotsc, T\})$
      \State $\bepsilon\sim\mathcal{N}(\bzero,\bI)$
      \State Take gradient descent step on
      \Statex $\quad \grad_\theta \left\| \bepsilon - \bepsilon_\theta(\sqrt{\alpha_t} \boldsymbol{\hat{x}} + \sqrt{1-\alpha_t}\boldsymbol{\epsilon}, t, \boldsymbol{y}) \right\|^2$
    \Until{converged}
  \end{algorithmic}
\end{algorithm}
\end{minipage}
\hfill
\begin{minipage}[t]{0.495\textwidth}
\begin{algorithm}[H]
  \caption{Inference} \label{alg:inference}
  \small
  \begin{algorithmic}[1]
    \vspace{.04in}
    \State $\boldsymbol{x} \sim q(\boldsymbol{x})$
    \State $\boldsymbol{\hat{y}} = f(\boldsymbol{x})$
    \For{$t=T_m, \dotsc, 1$}
      \State $\bz \sim \mathcal{N}(\bzero, \bI)$ if $t > 1$, else $\bz = \bzero$
      \State \begin{tiny}$\boldsymbol{\hat{y}}_{t-1} = \frac{1}{\sqrt{\alpha_t}}\left( \boldsymbol{\hat{y}}_t - \frac{1-\alpha_t}{\sqrt{1-\alpha_t}} \bepsilon_\theta(\boldsymbol{\hat{y}}_t, t, \boldsymbol{\hat{y}}) \right) + \sigma_t \bz$ \end{tiny}
    \EndFor
    \State \textbf{return} $\boldsymbol{\hat{y}}_0$
    \vspace{.07in}
  \end{algorithmic}
\end{algorithm}
\end{minipage}
\vspace{-4mm}
\end{figure}

\section{Experiments}
\subsection{Implementation Details}
The degradation model is trained under the default setting of CycleGAN~\cite{cycle}. We use the same noise estimation network and noise schedule in~\cite{DDPM,DDIM} and set $T$ to 1000 when training LED. We concatenate the conditional image and the corrupted image as the input to the diffusion model. All images are resized into 512 $\times$ 512. We train LED for 150 epochs with a batch size of 8. The learning rate is initialized as $1e-5$ and decays following a cosine policy. In the inference phase, we set $T_m$ to be smaller than $T$. For one-step enhancement, we set $T_m$ as 800, while for coarse-to-fine enhancement we set it as 200. In practice, we speed up the inference process using the reverse process proposed in DDIM~\cite{DDIM}.

\subsection{Datasets}
We train the proposed LED on the EyeQ dataset~\cite{eyeq}, which contains 28792 fundus images labeled with three quality grades (``Good'', ``Usable'' and ``Reject''). In this work, we select ``Good'' images as high-quality images and ``Usable'' images as low-quality ones from the official training set. We train the degradation model in an unpaired manner with this unpaired dataset. For training LED, we exclusively utilize the high-quality images.  

We evaluate the enhancement performance on the following four datasets, all of which are degraded by the degradation model in~\cite{CoFoNet} for a restoration purpose.

\noindent {\bf GAMMA}~\cite{gamma} contains 100 fundus images with glaucoma annotations, fovea location labels and optic disc/optic cup (OD/OC) segmentation masks.

\noindent {\bf DRIVE}~\cite{drive} contains 40 fundus images with vessel segmentation masks.

\noindent {\bf REFUGE}~\cite{refuge} contains 1200 fundus images with OD/OC segmentation masks. 

\noindent {\bf FIVES}~\cite{fives} contains 800 fundus images with labels of three eye diseases.

\begin{figure}[t]
    \centering
    \includegraphics[width=0.85\textwidth]{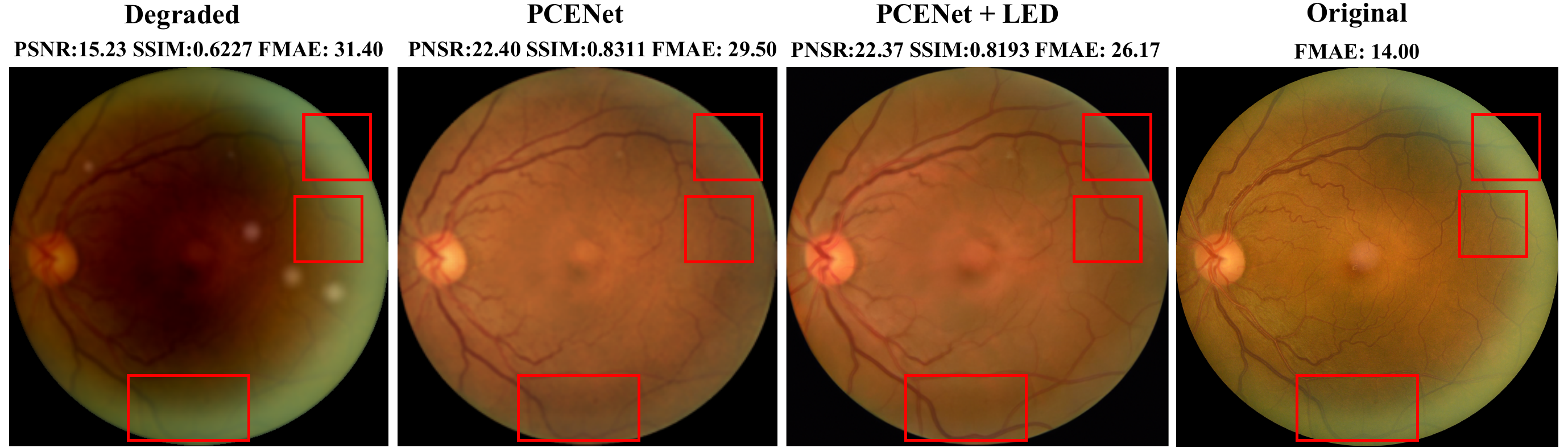}
    \vspace{0mm}
    \caption{From left to right: the degraded image, the coarse enhancement result from PCENet, the fine enhancement result from LED and the reference image. Although the PSNR and SSIM are lower, the enhancement from LED is significantly more visually realistic than that from PCENet and its FMAE for fovea localization is much lower.}
    \label{fig:psnr}
    \vspace{0mm}
  \end{figure}

\subsection{Evaluation Metrics}
Because the reference high-quality images are usually not available, high-quality and low-quality image pairs are usually obtained through synthesization. For this type of reference evaluation, various image restoration metrics such as PSNR and SSIM are widely used~\cite{psnr}. However, these metrics only focus on average low-level features. As shown in Fig. \ref{fig:psnr}, higher values of PSNR or SSIM do not necessarily indicate better preservation of structural details, as already pointed out elsewhere~\cite{srgan}.  To assess the fundus image enhancement performance with respect to structural details and clinically-relevant applications, we propose the following evaluation metrics: \\
\noindent {\bf FIQA}. Similar to~\cite{I-SECRET}, we train a ResNet101~\cite{resnet} on EyeQ for quality assessment and use the predictions to calculate a fundus image quality assessment score (FIQA). We evaluate on the original testing set of EyeQ labeled as ``Usable''. \\ 
\noindent {\bf OCSD}. We use the OC segmentation Dice (OCSD) as a metric of assessing the capability of preserving anatomical structures. We first train a U-Net~\cite{unet} on GAMMA and then evaluate on the testing set of the degraded REFUGE dataset. \\
\noindent {\bf VSD}. We use the vessel segmentation Dice (VSD) as another metric of assessing the capability of preserving anatomical structures. Iter-Net~\cite{iter} is used to segment vessels on the testing set of the degraded DRIVE dataset. \\ 
\noindent {\bf FCNR}. Inspired by the contrast noise ratio~\cite{cnr}, we propose a fundus contrast noise ratio (FCNR) to evaluate the contrast quality of the vessel areas. We first obtain a region of interest (ROI) $R$ using a disk-shaped dilation operation with a radius of 3 pixels based on the vessel segmentation mask. We formulate the vessel area as $V$ while the background area as $B$. The FCNR is defined as $\frac{| \mu_B - \mu_V |}{\sigma_R}$, where $\mu_B$ and $\mu_V$ are respectively the mean intensities of the background and vessel areas, and $\sigma_R$ is the intensity's standard deviation within the entire ROI area. FCNR is calculated on the testing set of the degraded DRIVE dataset. \\
\noindent {\bf FMAE}. We use a fovea detection network proposed in~\cite{joint} to assess the fovea location's mean absolute error (FMAE) on the degraded GAMMA dataset. \\
\noindent {\bf DRA}. We use the DR detection accuracy (DRA) as a metric for quantifying clinically-meaningful semantic enhancement. We utilize the DR detection network proposed in~\cite{lesion-dr} to discriminate DR on the degraded FIVES dataset.

% \usepackage{multirow}
% \usepackage{graphicx}
\begin{table}[t]
  \caption{Comparisons of the proposed LED with other SOTA methods.  \textbf{-} indicates that the metric is not applicable. The improvement induced by LED is marked in \bf{\textcolor{orange}{orange}}.}
  \centering
  \resizebox{1.0\textwidth}{!}{%
  \begin{tabular}{lclccccc}
  \toprule
  \multirow{2}{*}{Methods} & Non-reference &  & \multicolumn{5}{c}{Full-reference}    \\ \cline{2-2} \cline{4-8} 
   &
    FIQA $\uparrow$ &
     &
    OCSD $\uparrow$ &
    VSD $\uparrow$ &
    FCNR $\uparrow$ &
    FMSE $\downarrow$ &
    DRA $\uparrow$ \\ \hline
  Original                 & 20.58        &  & 80.45 & 81.61 & 35.05 & 9.24  & 75.60 \\
  Degraded                 & \textbf{-}            &  & 37.58 & 60.76 & 12.34 & 34.80 & 68.00 \\
  LED &
    65.97\bf{\textcolor{orange}{($+$45.39)}} &
     &
    54.26\bf{\textcolor{orange}{($+$16.68)}} &
    70.86\bf{\textcolor{orange}{($+$10.10)}} &
    17.79\bf{\textcolor{orange}{($+$5.45)}} &
    29.69\bf{\textcolor{orange}{($-$5.11)}} &
    69.87\bf{\textcolor{orange}{($+$1.87)}} \\ \hline
  CycleGAN~\cite{cycle}                 & 82.54        &  & 28.73 & 63.46 & 13.52 & 39.99 & 39.73 \\
  CycleGAN+LED &
    98.22\bf{\textcolor{orange}{($+$15.68)}} &
     &
    29.89\bf{\textcolor{orange}{($+$1.16)}} &
    66.14\bf{\textcolor{orange}{($+$2.68)}} &
    17.97\bf{\textcolor{orange}{($+$4.45)}}  &
    38.53\bf{\textcolor{orange}{($-$1.46)}} &
    64.67\bf{\textcolor{orange}{($+$24.94)}} \\ \hline
  CUT~\cite{cut}                      & 65.34        &  & 37.61 & 69.49 & 12.84 & 37.04 & 51.33 \\
  CUT+LED           & 87.01\bf{\textcolor{orange}{($+$21.67)}}       &  & 43.11\bf{\textcolor{orange}{($+$5.50)}} & 71.54\bf{\textcolor{orange}{($+$2.05)}} & 16.07\bf{\textcolor{orange}{($+$3.23)}} & 36.49\bf{\textcolor{orange}{($-$0.55)}} & 64.40\bf{\textcolor{orange}{($+$13.07)}} \\ \hline
  SCRNet~\cite{SCRNet}                  & 93.90        &  & 52.83 & 68.84 & 16.97 & 60.86 & 54.00 \\
  SCRNet+LED               & 98.55\bf{\textcolor{orange}{($+$4.65)}}        &  & 53.21\bf{\textcolor{orange}{($+$0.38)}} & 71.09\bf{\textcolor{orange}{($+$2.25)}} & 23.33\bf{\textcolor{orange}{($+$6.36)}} & 44.50\bf{\textcolor{orange}{($-$16.36)}} & 64.27\bf{\textcolor{orange}{($+$10.27)}} \\ \hline
  ArcNet~\cite{ArcNet}                  & 45.15        &  & 42.41 & 66.79 & 12.96 & 36.82 & 61.46 \\
  ArcNet+LED               & 71.48\bf{\textcolor{orange}{($+$26.33)}}        &  & 51.18\bf{\textcolor{orange}{($+$8.77)}} & 69.98\bf{\textcolor{orange}{($+$3.19)}} & 17.09\bf{\textcolor{orange}{($+$2.05)}} & 32.41\bf{\textcolor{orange}{($-$4.41)}} & 69.07\bf{\textcolor{orange}{($+$7.61)}} \\ \hline
  I-SECRET~\cite{I-SECRET}                 & 90.71        &  & 61.37 & 72.14 & 22.08 & 24.47 & 50.93 \\
  I-SECRET+LED             & 93.07\bf{\textcolor{orange}{($+$2.36)}}        &  & 66.53\bf{\textcolor{orange}{($+$5.16)}} & 73.84\bf{\textcolor{orange}{($+$1.70)}} & 27.93\bf{\textcolor{orange}{($+$5.85)}} & 23.24\bf{\textcolor{orange}{($-$1.23)}} & 63.07\bf{\textcolor{orange}{($+$12.14)}} \\ \hline
  PCENet~\cite{PCENet}                   & 73.72        &  & 72.23 & 69.17 & 18.08 & 18.43 & 49.07 \\
  PCENet+LED               & 91.68\bf{\textcolor{orange}{($+$17.96)}}        &  & 75.39\bf{\textcolor{orange}{($+$3.16)}} & 72.02\bf{\textcolor{orange}{($+$2.85)}} & 24.92\bf{\textcolor{orange}{($+$6.84)}} & 17.09\bf{\textcolor{orange}{($-$1.34)}} & 73.73\bf{\textcolor{orange}{($+$24.66)}} \\ \bottomrule
  \end{tabular}%
  }
  \label{tab:compare}
  \vspace{-1.5cm}

  \end{table}

\subsection{Performance Evaluations}
We conduct quantitative experiments to evaluate the performance of the proposed LED, as tabulated in Table~\ref{tab:compare}. We firstly evaluate the performance of the one-step LED enhancement. Directly employing LED highlights structural details and clinical semantics, resulting in comparatively higher VSD and DRA. Since LED is a probabilistic diffusion model, which explicitly models the distribution of high-quality images, it tends to preserve global details and avoid over-enhancement such as illumination. To further enhance global information, we test four SOTA fundus image enhancement methods (SRCNet, ARCNet, I-SECRET, and PCENet) and two widely used unpaired image translation methods (CycleGAN and CUT) as the coarse enhancement network. Significant performance boosts are observed in terms of all metrics. It is worth noting that all methods other than LED show decreased performance on DRA compared to the original degraded images, which are evaluated on a multi-disease dataset. A plausible reason is that all other methods fail to distinguish the lesions of different diseases, while LED preserves clinically-meaningful information during the diffusion process. 

\begin{figure}[t]
    \centering
    \includegraphics[width=0.85\textwidth]{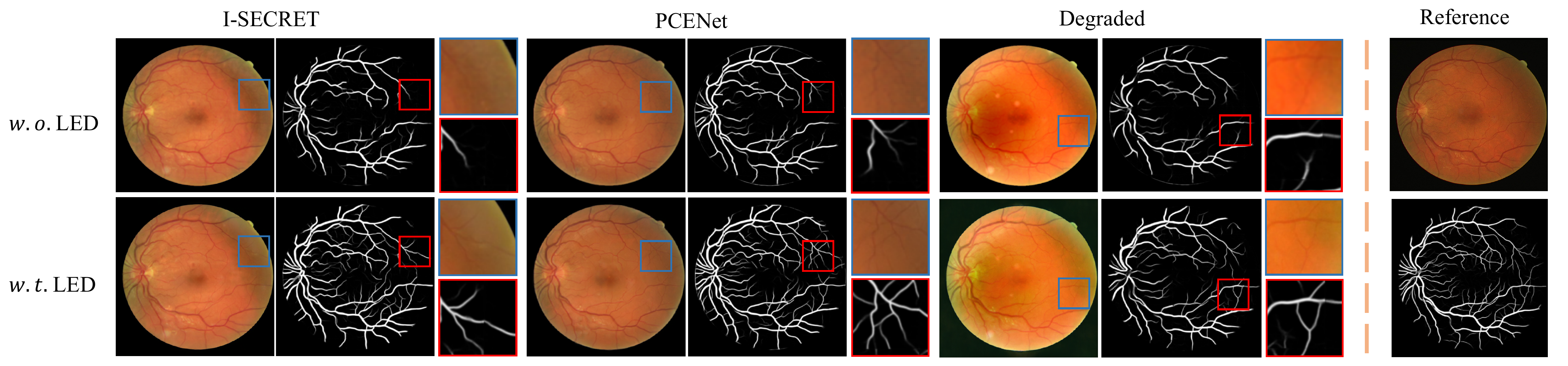}
    \vspace{-4mm}
    \caption{Visual comparisons on fundus image enhancement with respect to vessel segmentation. The first row shows the coarse enhancement results from PCENet and I-SECRET, as well as the images with no enhancement. The second row shows the corresponding fine enhancement results by incorporating our proposed LED. }
    \label{seg_compare}
    \vspace{-3mm}
    \end{figure}

\begin{figure}[t]
    \centering
    \includegraphics[width=0.85\textwidth]{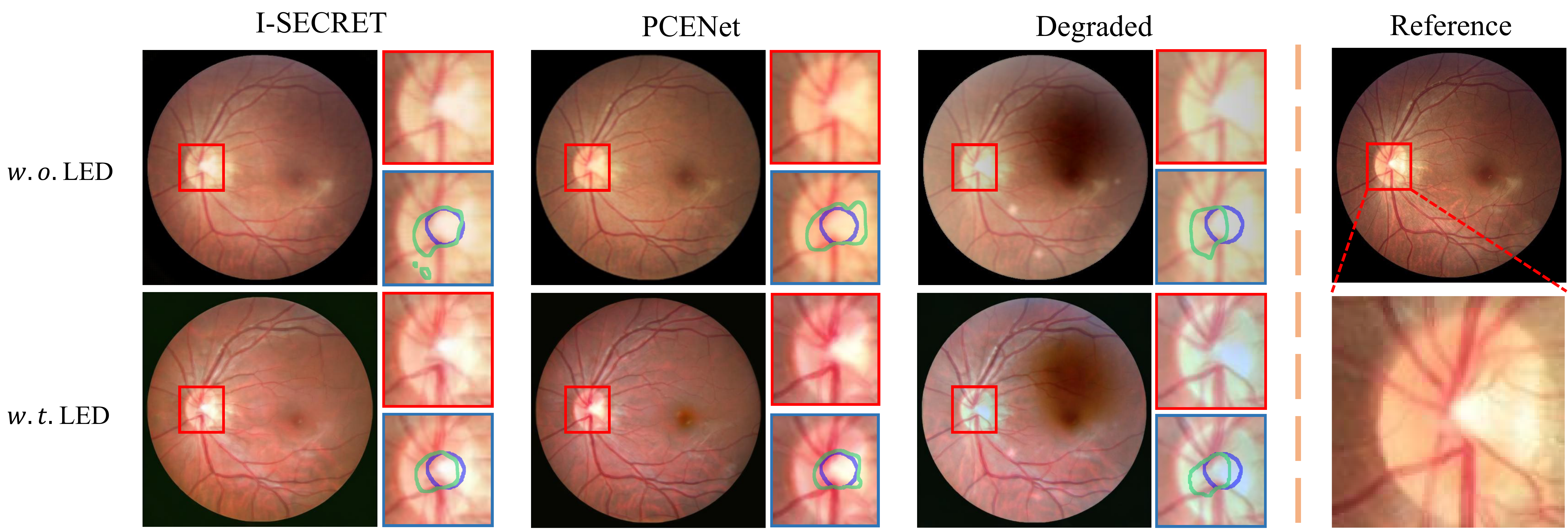}
    \vspace{-4mm}
    \caption{Visual comparisons on fundus image enhancement with respect to OC segmentation. Green contours indicate the segmentation results while blue contours indicate the ground truth. }
    \label{fig:oc}
    \vspace{0mm}
  \end{figure}

We present visual comparison results from two SOTA methods (PCENet and I-SECRET) in Fig.~\ref{seg_compare} and Fig.~\ref{fig:oc}. Apparently, LED significantly highlights vessels and OC with minimal information modification. These results suggest that, when combined with other SOTA methods, LED further emphasizes high-frequency details and boosts the corresponding segmentation performance of interest.

%Moreover, based on the diffusion process, LED is able to deliver controllable continuous enhancement. As shown in supplementary material, it is very flexible to adopt LED to control the enhancement level, which provides more explainability for the enhancement results.

%\begin{figure}[t]
    %\centering
    %\includegraphics[width=0.9\textwidth]{Continuous-v2.pdf}
    %\caption{By controlling the diffusion step, LED can %generate controllable continuous enhancement under %different setting of $T_m$.}
    %\label{continuous}
    %\vspace{-8mm}
  %\end{figure}

\section{Conclusion}
In this paper, we propose a novel diffusion model for fundus image enhancement, named LED, which aims to learn enhancement from learnable degradation. Since it models the distribution over high-quality images, LED can output realiable enhancement. We also propose a coarse-to-fine enhancement framework to further boost the performance on top of existing SOTA methods. Extensive experiments demonstrate the effectiveness of the proposed method, in terms of multiple clinically-meaning evaluation metrics and multiple fundus image datasets.

\section{Acknowledgements}
This study was supported by the Shenzhen Basic Research Program (JCYJ20190
809120205578); the National Natural Science Foundation of China (62071210); the Shenzhen Science and Technology Program (RCYX20210609103056042); the Shenzhen Basic Research Program (JCYJ2020092515384 7004); the Shenzhen Science and Technology Innovation Committee Program (KCXFZ2020122117340001).

\newpage

\begin{large} 
\textbf{\quad \quad \quad \quad \quad \quad \ Supplementary Material}
\end{large}

\begin{figure}[htb]
  \centering
  \includegraphics[width=1.0\textwidth]{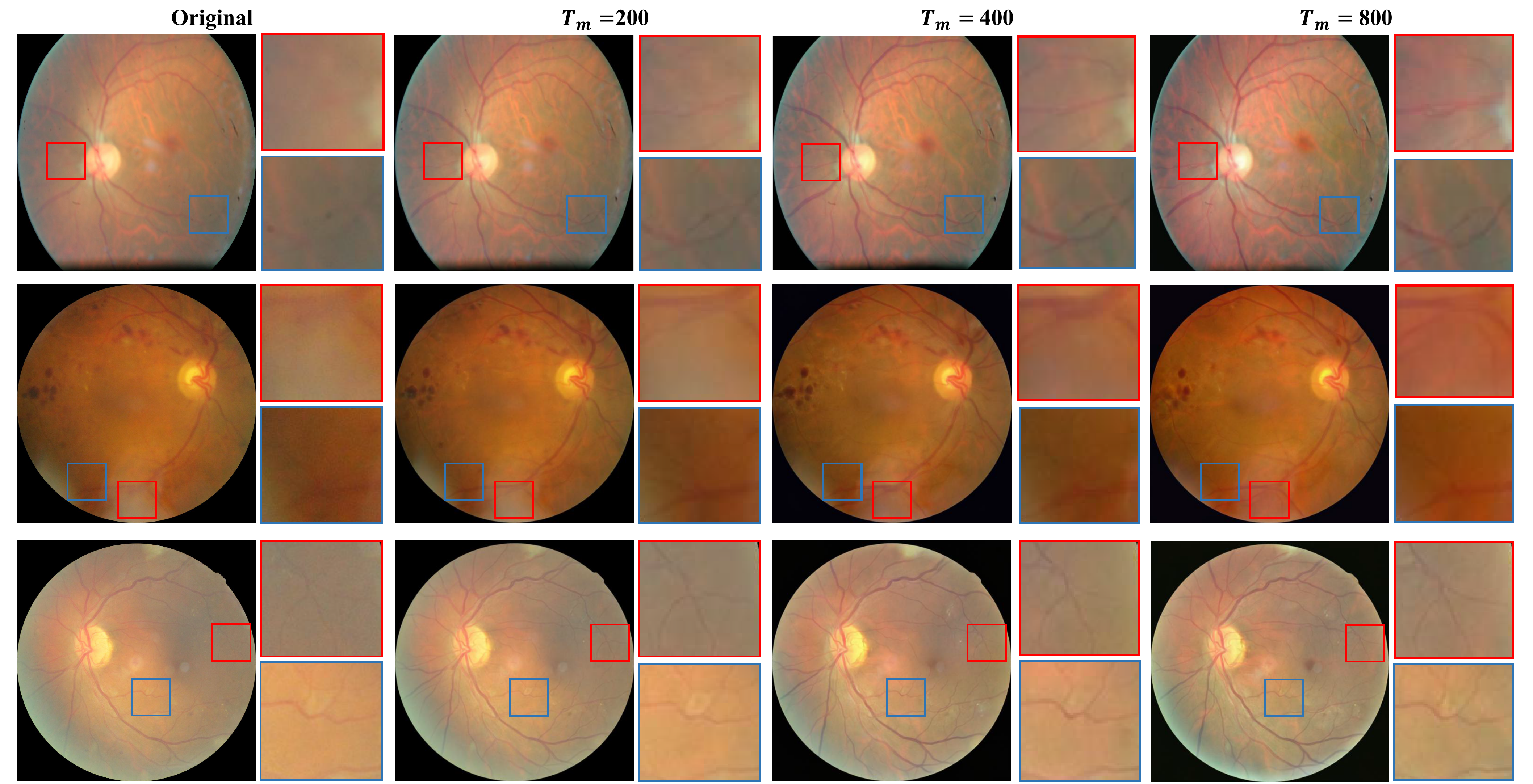}
  \vspace{-5mm}
  \caption{By controlling the diffusion steps, LED can deliver controllable and continuous enhancement results.}
  \label{continuous}
  \vspace{-5mm}
\end{figure}

% \usepackage{graphicx}
\begin{table}[htb]
  \centering
  \caption{Performance evaluations conducted on the degraded DRIVE testing set. Despite slight decreases in PSNR and SSIM, LED significantly boosts the vessel-relative metrics VSD and FCNR. The improvement induced by LED is marked in \textcolor{orange}{\bf{orange}} while the decrease is marked in \bf{\textcolor{RoyalBlue}{blue}}.}
  \resizebox{0.85\textwidth}{!}{%
  \begin{tabular}{lcccc}
  \toprule
  Method &
    PNSR  &
    SSIM &
    VSD &
    FCNR \\ \hline
  Degraded &
    15.85 &
    0.8050 &
    60.76 &
    12.34 \\
  Degraded+LED &
    16.48\bf{\textcolor{orange}{($+$0.63)}} &
    0.8071\bf{\textcolor{orange}{($+$0.0021)}} &
    70.86\bf{\textcolor{orange}{($+$10.10)}} &
    17.79\bf{\textcolor{orange}{($+$5.45)}} \\ \hline
  CycleGAN &
    17.20 &
    0.7710 &
    63.46 &
    13.52 \\
  CycleGAN+LED &
    17.18\bf{\textcolor{RoyalBlue}{($-$0.02)}} &
    0.7639\bf{\textcolor{RoyalBlue}{($-$0.0071)}} &
    66.14\bf{\textcolor{orange}{($+$2.68)}} &
    17.97\bf{\textcolor{orange}{($+$4.45)}} \\ \hline
  CUT &
    18.56 &
    0.8694 &
    69.49 &
    12.84 \\
  CUT+LED &
    18.67\bf{\textcolor{orange}{($+$0.11)}} &
    0.8621\bf{\textcolor{RoyalBlue}{($-$0.0073)}} &
    71.54\bf{\textcolor{orange}{($+$2.05)}} &
    16.07\bf{\textcolor{orange}{($+$3.23)}} \\ \hline
  SCRNet &
    22.55 &
    0.9058 &
    68.84 &
    16.97 \\
  SCRNet+LED &
    22.38\bf{\textcolor{RoyalBlue}{($-$0.17)}} &
    0.8906\bf{\textcolor{RoyalBlue}{($-$0.0152)}} &
    71.09\bf{\textcolor{orange}{($+$2.25)}} &
    23.33\bf{\textcolor{orange}{($+$6.36)}} \\ \hline
  ArcNet &
    17.24 &
    0.8275 &
    66.79 &
    12.96 \\
  AcrNet+LED &
    17.19\bf{\textcolor{RoyalBlue}{($-$0.05)}} &
    0.8079\bf{\textcolor{RoyalBlue}{($-$0.0196)}} &
    69.98\bf{\textcolor{orange}{($+$3.19)}} &
    17.09\bf{\textcolor{orange}{($+$4.13)}} \\ \hline
  I-SECRET &
    22.99 &
    0.9065 &
    72.14 &
    22.08 \\
  I-SECRET+LED &
    22.52\bf{\textcolor{RoyalBlue}{($-$0.47)}} &
    0.9007\bf{\textcolor{RoyalBlue}{($-$0.0058)}} &
    73.84\bf{\textcolor{orange}{($+$1.70)}} &
    27.93\bf{\textcolor{orange}{($+$5.85)}} \\ \hline
  PCENet &
    22.09 &
    0.9085 &
    69.17 &
    18.08 \\
  PCENet+LED &
    21.96\bf{\textcolor{RoyalBlue}{($-$0.13)}} &
    0.8943\bf{\textcolor{RoyalBlue}{($-$0.0412)}} &
    72.02\bf{\textcolor{orange}{($+$2.85)}} &
    24.92\bf{\textcolor{orange}{($+$6.84)}} \\ \bottomrule
  \end{tabular}%
  }
  \vspace{-10mm}
  \end{table}

  \begin{figure}[t]
    \centering
    \includegraphics[width=1.0\textwidth]{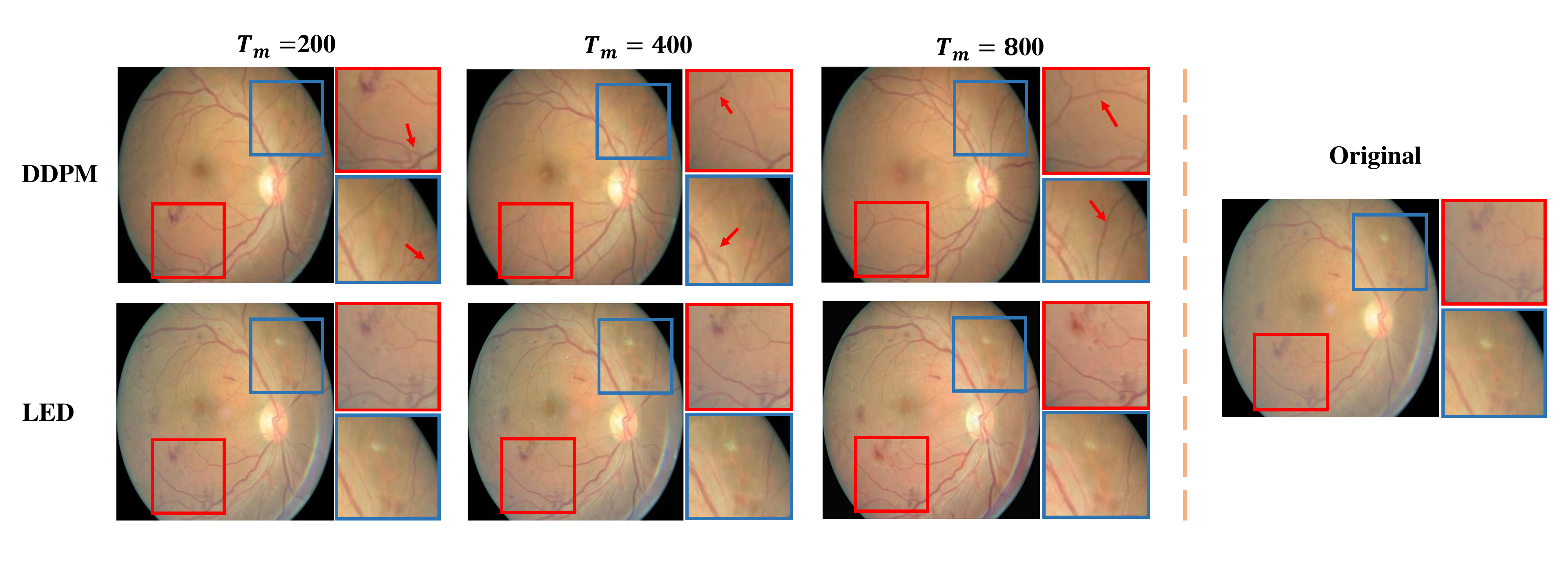}
    \vspace{-10mm}
    \caption{Visual comparisons between DDPM and LED. DDPM tends to modify structural and clinically relevant information, whereas LED not only preserves but also accentuates them.}
    \label{info}
    \vspace{-2mm}
  \end{figure}

  \begin{figure}[t]
    \centering
    \includegraphics[width=0.9\textwidth]{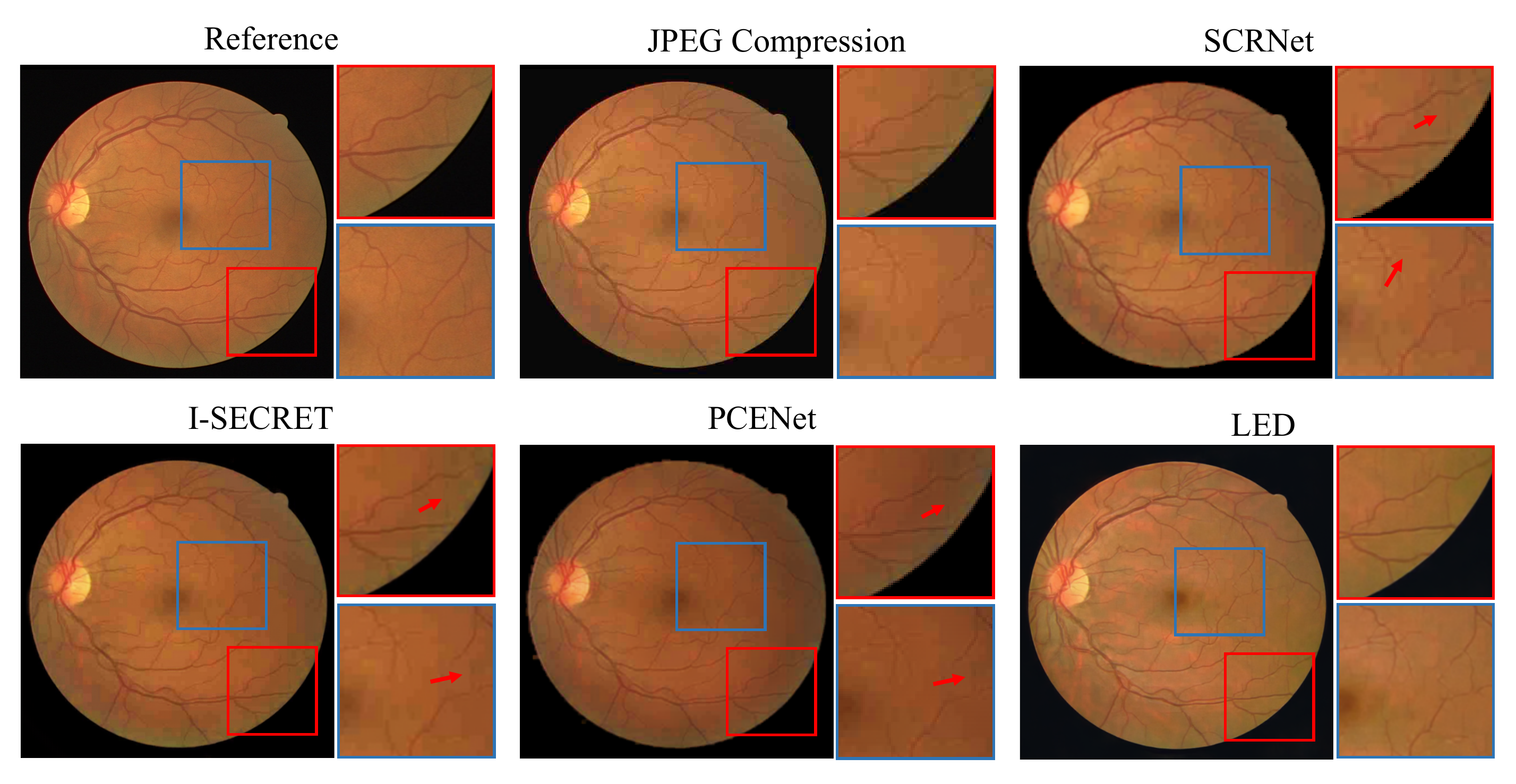}
    \vspace{-4mm}
    \caption{Visual comparisons between LED and other SOTA methods on out-of-distribution (OOD) low-quality images induced by the JPEG compression loss. LED generates more visually pleasing results than other methods.}
    \label{ood}
    \vspace{-2mm}
  \end{figure}

  \begin{figure}[b]
    \centering
    \includegraphics[width=0.9\textwidth]{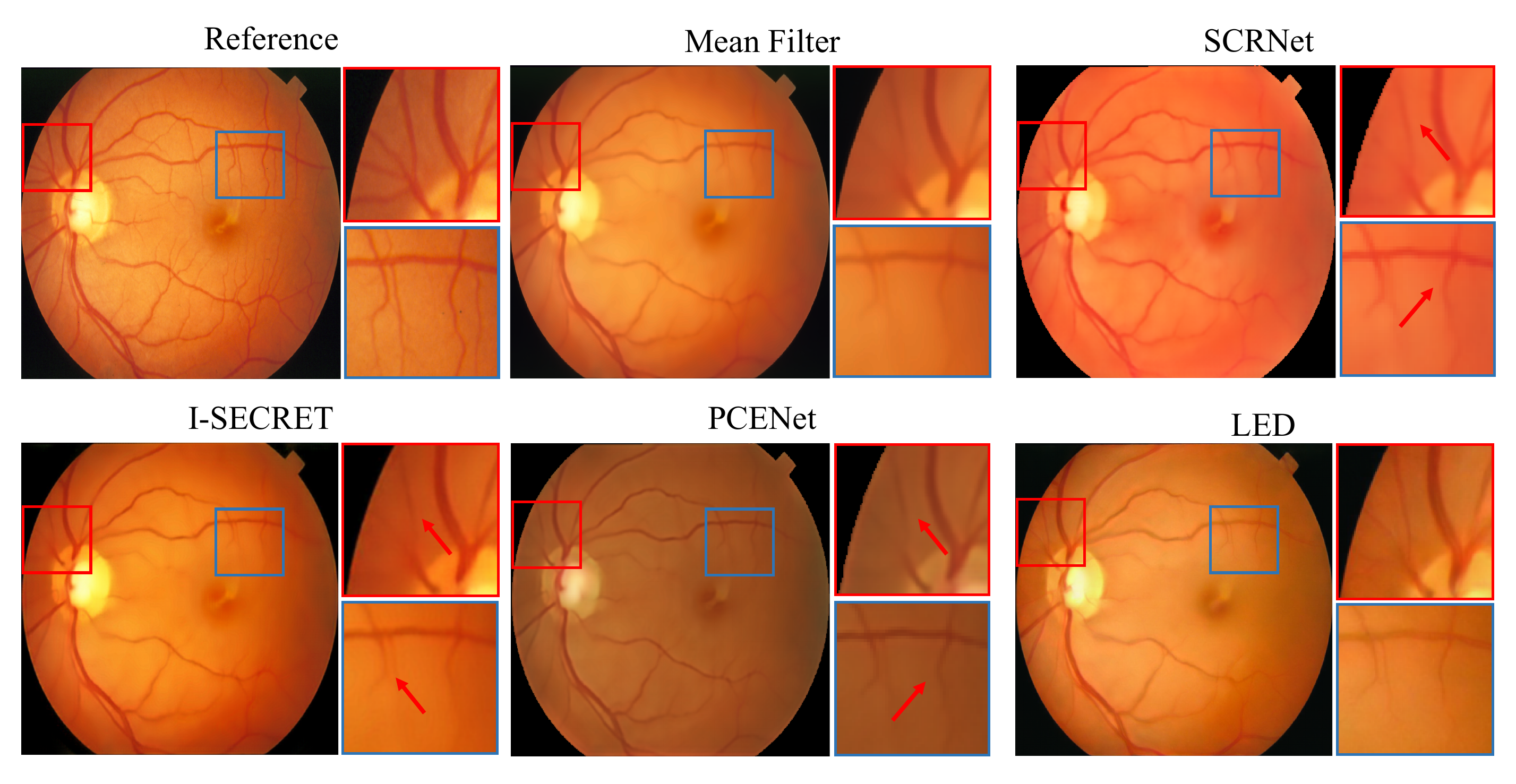}
    \vspace{-4mm}
    \caption{Visual comparisons between LED and other SOTA methods on out-of-distribution (OOD) low-quality images induced by mean filtering. LED highlights vessels significantly clearer than other methods.}
    \label{ood-mf}
  \end{figure}

\end{document}